\documentclass[amssymb]{revtex4}

\usepackage{graphicx}

\begin{document}

\title{Effect of node deleting on network structure}
\author{Ke Deng}
 \email{dengke@jsu.edu.cn}

\author{Heping Zhao}

\author{Dejun Li}

\affiliation{Department of Physics, Jishou University, Jishou, Hunan
416000, People's Republic of China}

\begin{abstract}
The ever-increasing knowledge to the structure of various real-world
networks has uncovered their complex multi-mechanism-governed
evolution processes. Therefore, a better understanding to the
structure and evolution of these networked complex systems requires
us to describe such processes in more detailed and realistic manner.
In this paper, we introduce a new type of network growth rule which
comprises of adding and deleting of nodes, and propose an evolving
network model to investigate the effect of node deleting on network
structure. It is found that, with the introduction of node deleting,
network structure is significantly transformed. In particular,
degree distribution of the network undergoes a transition from
scale-free to exponential forms as the intensity of node deleting
increases. At the same time, nontrivial disassortative degree
correlation develops spontaneously as a natural result of network
evolution in the model. We also demonstrate that node deleting
introduced in the model does not destroy the connectedness of a
growing network so long as the increasing rate of edges is not
excessively small. In addition, it is found that node deleting will
weaken but not eliminate the small-world effect of a growing
network, and generally it will decrease the clustering coefficient
in a network.
\end{abstract}

%\pacs{87.23.Kg, 89.75.Fb, 05.10.-a, 89.75.Hc}
%\keywords{node-deleting, structure, complex networks}
 \maketitle

\section{INTRODUCTION}
Network structure is of great importance in the topological
characterization of complex systems in reality. Actually, these
networked complex systems have been found to share some common
structural characteristics, such as the small-world properties, the
power-law degree distribution, the degree correlation, and so on
\cite{1,a1,2}. In the theoretical description of these findings, the
Watts-Strogatz (WS) model \cite{3} provides a simple way to generate
networks with the small-world properties. Barab\'{a}si and Albert
(BA) \cite{4}, with a somewhat different aim, proposed an evolving
network model to explain the origin of power-law degree
distribution. In this model, by considering two fundamental
mechanisms: growth and preferential attachment (PA), power-law
degree distribution emerges naturally from network evolution. Based
on the framework of BA model, many other mechanisms were introduced
into network evolution to reproduce some more complex observed
network structures \cite{5,6,7,8,9,10,11,12,13,14,a2,a5}, such as
the degree distribution of broad scale and single scale \cite{5}, as
well as the degree correlation \cite{a5}. These further studies show
that real networked systems may undergo a very complex evolution
process governed by multiple mechanisms on which the occurrence of
network structures depends. Therefore, to get a better understanding
of the structure and evolution of complex networks, describing such
processes in more detailed and realistic manner is necessary.

In the BA's framework, the growing nature of real-world networks is
captured by a BA-type growth rule. According to this rule, one node
is added into the network at each time step, intending to mimic the
growing process of real systems. This rule gives an explicit
description to the real-network' growing process which, however, can
in fact be much more complex. One fact is that in many real growing
networks, there are constant adding of new elements, but accompanied
by permanent removal of old elements (deletion of nodes)
\cite{28,15,16,17,a3,a4}. Take the food webs for a example: there
are both additions and losses of nodes (species) at ecological and
evolutionary time scales by means of immigration, emigration,
speciation, and extinction \cite{28}. Likewise, for Internet and the
World Wide Web (WWW), node-deleting is reported experimentally in
spit of their rapid expansion of size \cite{15,16,17,a3,a4}. In the
Internet's Autonomous Systems (ASs) map case, a node is an AS and a
link is a relationship between two ASs. An AS adding means a new
Internet Service Provider (ISP) or a large institution with multiple
stub networks joins the Internet. An AS deleting happens due to the
permanent shutdown of the corresponding AS as it is, for example,
out of business. Investigations of the evolution of real Internet
maps from 1997 to 2000 verified such network mechanism
\cite{15,16,17}. The same is for the evolution of WWW, in which the
deletions of invalid web pages are also frequently discovered
\cite{a3,a4}. In most cases, the deletion of a node is also
accompanied with the removal of all edges once attached to it. These
facts justify the investigation of node-deletion's influence on
network structure. In this paper, we introduce a new type of network
growth rule which comprises of adding and deleting of nodes, and
propose an evolving network model to investigate the effect of node
deleting on the network structure. Before now several authors have
proposed some models on node removal in networks, such as AJB
networks in which a portion nodes are simultaneously removed from
the network \cite{18}, and also the decaying \cite{19} and mortal
\cite{20} networks, which concerns networks' scaling property and
critical behavior respectively. Sarshar \textit{et al} \cite{21}
investigated the \textit{ad hoc} network with node removal, focusing
on the compensatory process to preserve true scale-free state. They
are different from present work, in which node deleting is treated
as an ubiquitous mechanism accompanied with the evolution of
real-world networks.

This paper is organized as follows. In Section~\ref{sec:model}, an
evolving network model taking account of the effect of node deleting
is introduced which reduces to a generalized BA model when the
effect of node deleting vanishes. Then the effect of node deleting
on network structure are investigated in five aspects: degree
distribution (Section~\ref{sec:pk}), degree correlation
(Section~\ref{sec:r}), size of giant component
(Section~\ref{sec:s}), average distance between nodes
(Section~\ref{sec:l}) and clustering (Section~\ref{sec:c}). Finally,
Section~\ref{sec:conclusion} presents a brief summary.

\section{\label{sec:model}THE MODEL}
We consider the following model. In the initial state, the network
has $m_0$ isolated nodes. At each time step, either a new node is
added into the network with probability $P_a$ or a randomly chosen
old node is deleted from the network with probability $P_d=1-P_a$,
where $P_a$ is an adjustable parameter. When a new node is added to
the network, it connects to $m$ ($m\leqslant m_0$) existing node in
the network according to the preferential probability introduced in
the BA model \cite{4}, which reads
\begin{eqnarray}
\Pi_\alpha=\frac{k_\alpha+1}{\sum_\beta(k_\beta+1)}
\label{eq:1}
\end{eqnarray}
where $k_\alpha$ is the degree of node $\alpha$. When an old node is
deleted from the network, edges once attached to it are removed as
well. In the model, $P_a$ is varied in the range of $0.5<P_a\leq1$,
since in the case of $P_a\leqslant 0.5$ the network can not grow. In
order to give a chance for isolated nodes to receive a new edge, we
choose preferential probability $\Pi_\alpha$ proportional to
$k_\alpha+1$ \cite{6}. Note that when $P_a=1$, our model reduces to
a generalized BA model \cite{22}.

To get a general knowledge to the effect of node deleting on network
structure, firstly, a simple analysis to the surviving probability
$D(i,t)$ is helpful. Here, $D(i,t)$ is defined as the probability
that a node is added into the network at time step $i$, and this
node (the $i$th node) has not been deleted until time step $t$,
where $t\geqslant i$. Supposing that a node-adding event happens at
time step $i^{'}$, and the probability that the $i'$th node has not
been deleted until time step $t$ is denoted as $D'(i',t)$. Then, due
to the independence of events happened at each time step, it is easy
to verify that $D'(i',t+1)=D'(i',t)[1-(1-P_a)/N(t)]$ with
$D'(i',i')=1$, where $N(t)=(2P_a-1)t$ is the number of nodes in the
network at moment $t$ (in the limit of large $t$). In the continuous
limit, we obtain
\begin{eqnarray}
\frac{\partial D'(i',t)}{\partial t}
=-\frac{(1-P_a)}{(2P_a-1)t}D'(i',t),
\end{eqnarray}
which yields
\begin{eqnarray}
D'(i',t)=\left( \frac{t}{i'}\right)^{-(1-P_a)/(2P_a-1)}.
\end{eqnarray}
Thus to get the $D(i,t)$ we should multiply $D'(i',t)$ with $P_a$,
i.e.
\begin{eqnarray}
D(i,t)=P_a\left( \frac{t}{i}\right)^{-(1-P_a)/(2P_a-1)}.
\label{eq:4}
\end{eqnarray}
One can easily find that $D(i,t)$ decreases rapidly as $t$ increases
and/or as $i$ decreases provided $0.5<P_a<1$. It is well known that
highly connected nodes, or hubs, play very important roles in the
structural and functional properties of growing networks
\cite{1,a1,2}. The formation of hubs needs a long time to gain a
large number of connections. As a consequence, according to
Eq.~(\ref{eq:4}), a large portion of potential hubs are deleted
during the network evolution. Thus it can be expected that the
introduction of node deleting has nontrivial effects on network
structure. In the following we show how network structure can be
effected by the node deleting introduced in present model.

\section{\label{sec:pk}DEGREE DISTRIBUTION}
The degree distribution $p(k)$, which gives the probability that a
node in the network possesses $k$ edges, is a very important
quantity to characterize network structure. In fact, $p(k)$ has been
suggested to be used as the first criteria to classify real-world
networks \cite{5}. Therefore it is necessary to investigate the
effect of node deleting on the degree distribution of networks
firstly. Now we adopt the continuous approach \cite{23} to give a
qualitative analysis of $p(k)$ for our model with slight node
deletion (i.e., when $P_d$ is very small). Supposing that there is a
node added into the network at time step $i'$, and this node is
still in the network at time $t$, let $k(i',t)$ be the degree of the
$i'$th node at time $t$, where $t\geqslant i'$. Then the increasing
rate of $k(i',t)$ is
\begin{eqnarray}
\frac{\partial k(i',t)}{\partial t} =P_am
\frac{k(i',t)+1}{S(t)}-(1-P_a) \frac{k(i',t)}{N(t)},
\label{eq:5}
\end{eqnarray}
where
\begin{eqnarray}
S(t)=\sum_{i^{'}}D'(i',t)[k(i',t)+1]
\label{eq:6}
\end{eqnarray}
and the $\sum_{i^{'}}$ denotes the sum of all $i'$ during the time
step between $0$ and $t$. It is easy to verify that the first term
in Eq.~(\ref{eq:5}) is the increasing number of links of the $i'$th
node due to the preferential attachment made by the newly added
node. The second term in Eq.~(\ref{eq:5}) accounts for the losing of
a link of the $i'$th node during the process of node deletion, which
happened with the probability $k(i',t)/N(t)$.

Firstly we solve for the $S(t)$ and get
\begin{eqnarray}
S(t)=\left(2P_a-1\right)\left(2P_am+1\right)t
\label{eq:7}
\end{eqnarray}
(see the Appendix for details). Inserting Eq.~(\ref{eq:7}) back into
Eq.~(\ref{eq:5}), one gets
\begin{eqnarray}
\frac{\partial k(i',t)}{\partial t} =\frac{Ak(i',t)+B}{t},
\label{eq:8}
\end{eqnarray}
where
\begin{eqnarray}
A=\frac{2P_a^{2}m-P_am+P_a-1}{(2P_a-1)(2P_am+1)}
\label{eq:9}
\end{eqnarray}
and
\begin{eqnarray}
B=\frac{P_am}{(2P_a-1)(2P_am+1)}.
\label{eq:10}
\end{eqnarray}
When $Ak+B>0$, the solution of Eq.~(\ref{eq:8}) is
\begin{eqnarray}
k(i',t)=\frac{1}{A}\left[(Am+B)\left(\frac{t}{i'}\right)^{A}-B\right].
\label{eq:11}
\end{eqnarray}

Now, to get the probability $p(k,t)$ that a randomly selected node
at time $t$ will have degree $k$, we need to calculate the expected
number of nodes $N_k(t)$ with degree $k$ at time $t$. Then the
$p(k,t)$ can be obtained from $p(k,t)=N_k(t)/N(t)$, where $N(t)$ is
the total number of nodes at time $t$. Let $I_k(t)$ represent the
set of all possible nodes with degree $k$ at time $t$, then one gets
\begin{eqnarray}
p(k,t)=\frac{N_k(t)}{N(t)}= \frac{1}{N(t)}\sum_{i\in I_k(t)} D(i,t).
\label{eq:12}
\end{eqnarray}

In the continuous-time approach, the number of nodes in $I_k(t)$ is
the number of $i$'s for which $k\leqslant k(i,t) \leqslant k+1$, and
it is approximated to $|\partial k(i,t) /
\partial i|^{-1}_{i=i_{k}}$, where $i_k$ is the solution of the
equation $k(i,t)=k$. To proceed with our analysis, now we make the
approximation that all nodes in $I_k(t)$ have the same surviving
probability $D(i_k,t)$ \footnote{It seems that this is not a very
good approximation, since investigations indicate that values of
$\left(\partial D(i,t)/\partial i\right)|_{i=i_{k}}$ are large and
increase rapidly with the decrease of $P_a$. Thus the analysis here
is a qualitative one and only suit for the condition of slight node
deletion in the model.}. Under this mean-field approximation,
Eq.~(\ref{eq:12}) can be written as
\begin{eqnarray}
p(k,t)=\frac{1}{N(t)}D(i_k,t)\left|\frac{\partial k(i,t)} {\partial
i}\right|^{-1}_{i=i_{k}}.
\label{eq:13}
\end{eqnarray}
From Eq.~(\ref{eq:11}), we obtain
\begin{eqnarray}
i_k=\left( \frac{Ak+B}{Am+B}\right)^{-1/A}t.
\label{eq:14}
\end{eqnarray}
then
\begin{equation}
\left|\frac{\partial k(i,t)} {\partial
i}\right|^{-1}_{i=i_{k}}=\left(Am+B\right)^{1/A}t\left(Ak+B\right)^{-(A+1)/A}.
\label{eq:15}
\end{equation}
Inserting Eq.~(\ref{eq:14}) back into Eq.~(\ref{eq:4}) we get
\begin{eqnarray}
D(i_k,t)=P_a\left(\frac{Ak+B}{Am+B}\right)^{(A-B)/A}
\label{eq:16}
\end{eqnarray}
Inserting Eqs.~(\ref{eq:15}) and ~(\ref{eq:16}) into
Eq.~(\ref{eq:13}), and noting that $N(t)=(2P_a-1)t$, we get
\begin{widetext}
\begin{equation}
p(k,t)=\frac{P_a}{2P_a-1}\left(Am+B\right)^{(B-A+1)/A}\left(Ak+B\right)^{-(B+1)/A},
\label{eq:17}
\end{equation}
\end{widetext}
which is a generalized power-law form with the exponent
\begin{equation}
\gamma=\frac{B+1}{A}=2+\frac{P_am+1}{2P_a^{2}m-P_am+P_a-1}.
\label{eq:18}
\end{equation}

\begin{figure}
\includegraphics[scale=0.8]{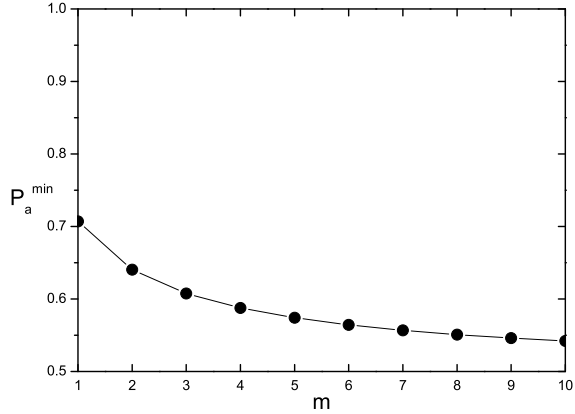}
\caption{\label{pam} $P_a^{min}$ [defined in Eq.~(\ref{eq:20})] as a
function of $m$.}
\end{figure}

We point out again that equation (\ref{eq:11}) is only valid when
$Ak+B>0$, which translates into $A>0$, i.e.
\begin{eqnarray}
2P_a^{2}m-P_am+P_a-1>0.
\label{eq:19}
\end{eqnarray}
Considering that $P_a>0.5$, Eq.~(\ref{eq:19}) is satisfied when
\begin{eqnarray}
P_a>P_a^{min}=\frac{(m-1)+\sqrt{m^{2}+6m+1}}{4m}.
\label{eq:20}
\end{eqnarray}
In Fig.~\ref{pam}, we plot $P_a^{min}$ as a function of $m$. One can
see from Fig.~\ref{pam} that the curve divides our model into two
regimes. $(i)$ $P_a>P_a^{min}$: in this case $Ak+B>0$ and equation
(\ref{eq:11}) is valid. Thus, the degree distribution of the network
$p(k)$ exhibits a generalized power-law form. $(ii)$
$P_a>P_a^{min}$: In this case $Ak+B>0$ can not be always satisfied
and equation (\ref{eq:11}) is not valid. Therefore, our continuous
approach fails to predict the behavior of $p(k)$, and we will
investigate it with numerical simulations. The $P_a^{min}(m)$, as
one can find from Fig.~\ref{pam}, decreases with the increase of
$m$.

\begin{figure}
\includegraphics[scale=0.8]{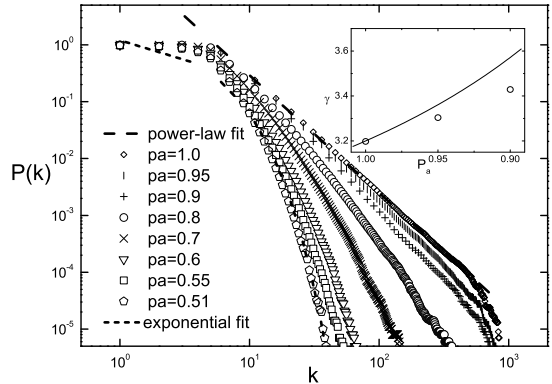}
\caption{\label{pk} Cumulative degree distribution $P(k)$ for
networks with system size $N=100000$ and different values of $P_a$,
in logarithmic scales. The dash line is power-law fit for $P_a=1$.
The solid line is the exponential fit for $P_a=0.51$. In the
simulation, we set $m_0=m=5$ and each distribution is based on $10$
independent realizations. Inset plots the power-law exponential
$\gamma$ as a function of $P_a$. The continuous curve is according
to the analytic result of Eq.~(\ref{eq:18}), and circles to the
simulation results.}
\end{figure}

In the power-law regime [$P_a>P_a^{min}(m)$], the behavior of $p(k)$
is predicted by Eqs.~(\ref{eq:17}) and~(\ref{eq:18}), which are
obtained using a mean-field approximation [Eq.~(\ref{eq:13})]. One
can easily verify that such approximation is only exact when
$P_a=1$, in which case Eq.~(\ref{eq:18}) turns into $\gamma=3+1/m$,
in good agreement with the results obtained from generalized BA
model studied in Ref \cite{22}. If $P_a^{min}(m)<P_a<1$,
Eqs.~(\ref{eq:17}) and~(\ref{eq:18}) still give qualitative
predictions for the model: with slight node deletion, $p(k)$ of the
network is still power-law, and the exponential $\gamma$ increases
with the decrease of $P_a$ (inset of Fig.~\ref{pk}).

In remaining regime [$P_a<P_a^{min}(m)$], the limiting case is
$P_a\rightarrow0.5$, in which the growth of network is suppressed (a
very slowly growing one). Similar non-growing networks have been
studied, for example, for the Model B in Ref\cite{24}, and the
degree distribution has the exponential form. Here we conjecture
that, in this regime, $p(k)$ of our model crossovers to an
exponential form, which is verified by the numerical simulation
results below.

Now we verify the above analysis with numerical simulations. In
Fig.~\ref{pk}, we give the cumulative degree distributions $P(k)$
\cite{2} of the networks with different $P_a$. As $P_a$ gradually
decreases from $1$ to $0.5$, Fig.~\ref{pk} shows an interesting
transition process which can be roughly divided into three stages.
$(1)$ $0.9\leqslant P_a\leqslant 1$: In this stage, the model works
in the power-law regime and the power-law exponent $\gamma$
increases as $P_a$ decreases. Inset of Fig.~\ref{pk} gives the
comparison between the value of $\gamma$ predicted by
Eq.~(\ref{eq:18}) and the one obtained from numerical simulations.
One sees that the theory and the simulation results are in perfect
agreement for $P_a=1$. As $P_a$ decreases, however, the agreement is
only qualitative and the deviation between theory and simulation
becomes more and more obvious. As we have mentioned above, such
increasing deviation is due to the mean-field approximation used in
the analysis. These results tell us that slight node deletion does
not cause deviation of the network from scale-free state, but only
increases its power-law exponent. Such robustness of power-low
$p(k)$ revealed here gives an explanation to the ubiquity of
scale-free networks in reality. It should be noted that a very
similar robustness has also been found in the study of network
resilience, where simultaneously deleting of a portion of nodes was
taken into account in static scale-free networks \cite{18}. $(2)$
$0.5<P_a\leqslant 0.6$: In this stage, the model works in the regime
of $P_a<P_a^{min}(m)$. As one sees from Fig.~\ref{pk}, $P(k)$ of the
network behaviors exponentially. This result indicates that with
manifest node deletion, the network will deviate from scale-free
state and become exponential. $(3)$ $0.6<P_a<0.9$: In this stage, a
crossover of the model from the power-law regime to the exponential
regime is found, in which the $P(k)$ is no longer pure scale-free
but truncated by an exponential tail. As one can see, the truncation
in $P(k)$ increases as $P_a$ decreases.

Besides the power-law degree distribution, it is now known that
$p(k)$ in real world may deviate from a pure power-law form
\cite{25,26,27,28,29}. According to the extent of deviation, $p(k)$
of real systems has been classified into three groups \cite{5}:
scale-free (pure power-law), broad scale (power-law with a
truncation), and single scale (exponential). Many mechanisms, such
as aging \cite{5,7,8}, cost \cite{5}, and information filtering
\cite{9}, have been introduced into network growth to explain these
distributions. Here, the results of Fig.~\ref{pk} indicate that a
modified version of growth rule can lead to all the three kinds of
$p(k)$ in reality, and it provides another explanation for the
origin of the diversity of degree distribution in real-world: such
diversity may be a natural result of network growth.

\section{\label{sec:r}DEGREE CORRELATION}
It has been recently realized that, besides the degree distribution,
structure of real networks are also characterized by degree
correlations \cite{15,30,31,32,33}. This translates into the fact
that degrees at the end of any given edge in real networks are not
usually independent, but are correlated with one another, either
positively or negatively. A network in which the degrees of adjacent
nodes are positively (negatively) correlated is said to show
assortative (disassortative) mixing by degree. An interesting
observation emerging from the comparing of real networks of
different types is that most social networks appear to be
assortatively mixed, whereas most technological and biological
networks appear to be disassortative. The level of degree
correlation can be quantified by the assortativity coefficient $r$
lying in the range $-1\leqslant r\leqslant1$, which can be written
as
\begin{equation}
r=\frac{M^{-1}\sum_{i}j_{i}k_{i}-\left[M^{-1}\sum_{i}\frac{1}{2}\left(j_{i}+k_{i}\right)\right]^{2}}{M^{-1}\sum_{i}\frac{1}{2}\left(j_{i}^{2}+k_{i}^{2}\right)-\left[M^{-1}\sum_{i}\frac{1}{2}\left(j_{i}+k_{i}\right)\right]^{2}}
\end{equation}
for practical evaluation on an observed network, where $j_{i}$,
$k_{i}$ are the degrees of the vertices at the ends of the $i$th
edge, with $i=1,\ldots,M$ \cite{30}. This formula gives $r>0 (r<0)$
when the corresponding network is positively (negatively)
correlated, and $r=0$ when there is no correlation \footnote{Another
way to represent degree correlation is to calculate the mean degree
of the nearest neighbors of a vertex as a function of the degree $k$
of that vertex. Although such way is explicit to characterize degree
correlation for highly heterogeneously organized networks, for less
heterogeneous networks (this is the case in the proposed model when
the intensity of node deleting increases, see Fig.~\ref{pk}), it may
be very nosy and difficult to interpret. So here we adopt the
assortativity coefficient $r$ to characterize degree correlation in
the model.}.

\begin{figure}
\includegraphics[scale=0.8]{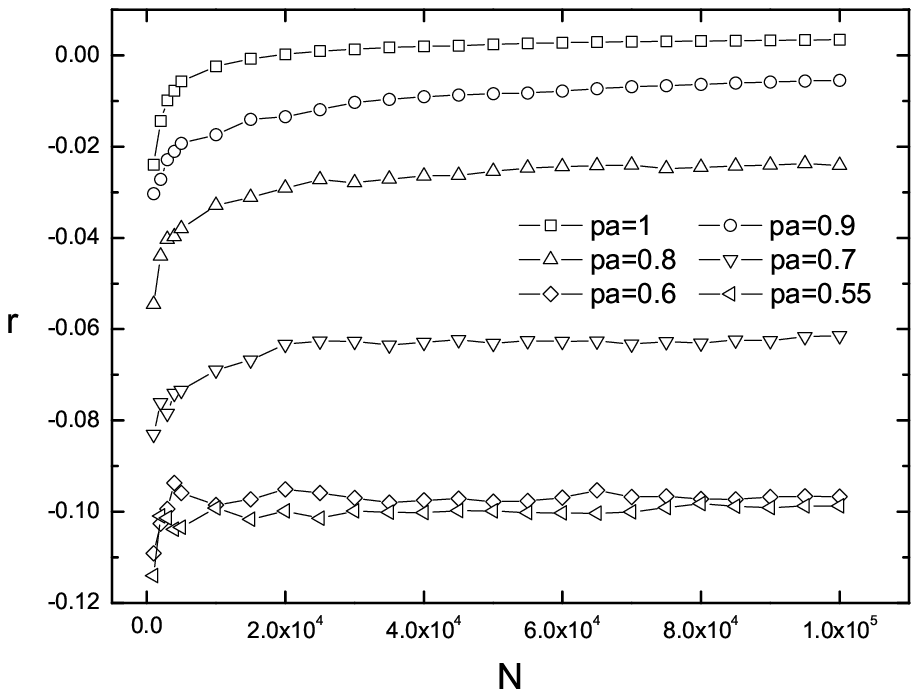}
\caption{\label{r} Assortativity coefficient $r$ plotted with
network size $N$, for different $P_a$ in the model. In the
simulation, $m_0=m=5$. Result of each curve is based on $10$
independent realizations.}
\end{figure}

\begin{figure}
\includegraphics[scale=0.8]{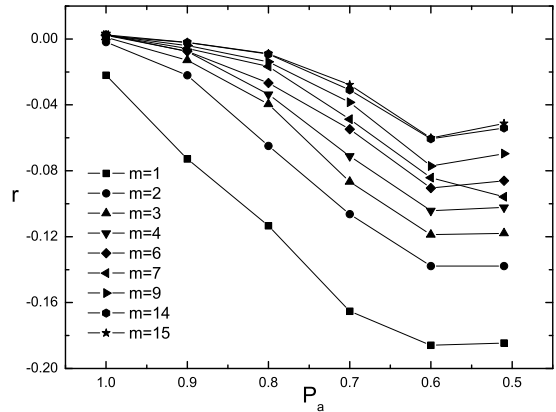}
\caption{\label{r-m} Assortativity coefficient $r$ as a function of
$P_a$, for different $m$ in the model. In the simulation, $N=40000$.
Result of each curve is based on $10$ independent realizations.}
\end{figure}

\begin{figure}
\includegraphics[scale=0.8]{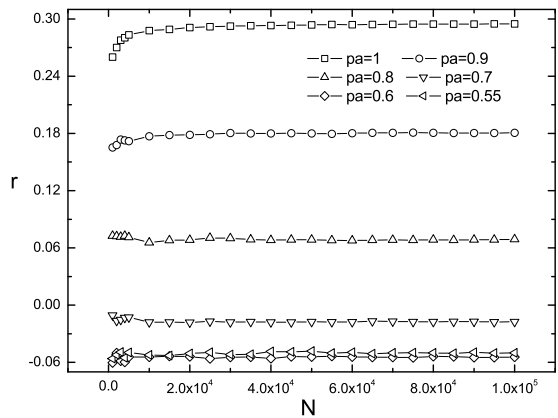}
\caption{\label{r-nopa} Assortativity coefficient $r$ plotted with
network size $N$, for different $P_a$ in the randomly growing
network model. In the simulation, $m_0=m=5$ and each curve is based
on $10$ independent realizations.}
\end{figure}

\begin{figure}
\includegraphics[scale=0.8]{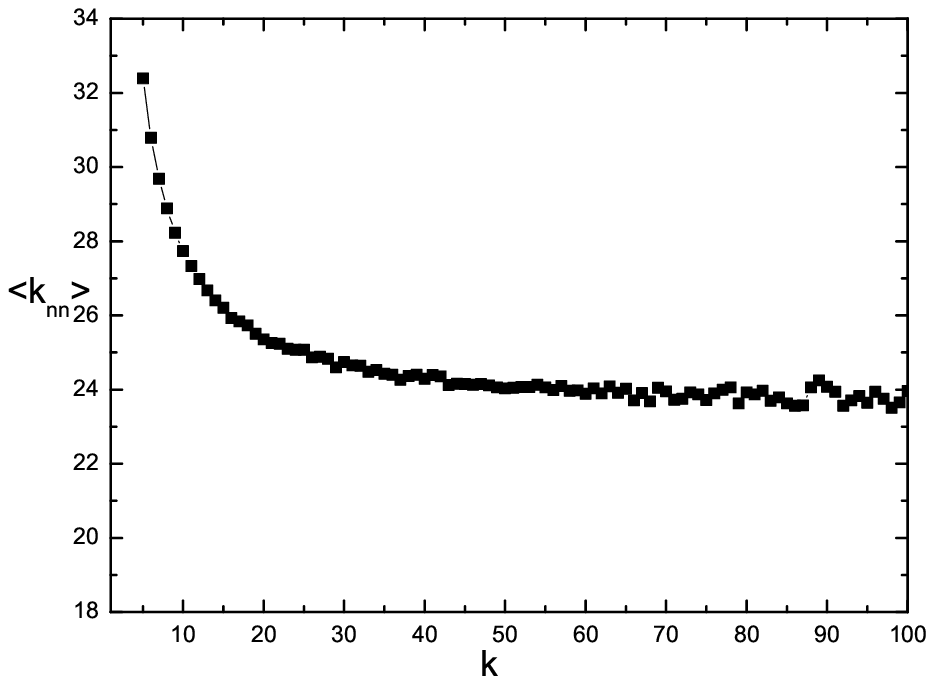}
\caption{\label{aknnba} Average degree of the nearest neighbor as a
function of $k$ for the BA model. In the simulation, $N=10000$ and
$m=m_0=5$. Result of each curve is based on $1000$ independent
realizations.}
\end{figure}

Recently, Maslov \textit{et al} \cite{34} and Park \textit{et al}
\cite{35} have proposed a possible explanation for the origin of
such correlation. They show for a network the restriction that there
is at most one edge between any pair of nodes induces negative
degree correlations. This restriction seems to be an universal
mechanism (indeed, there is no double edges in most real networks),
therefore, the authors of Ref. \cite{35} conjecture that
disassortativity by degree is the normal state of affairs for a
network. Although only a part of the measured correlation can be
explained in the way of Ref. \cite{35}, this universal mechanism
does give a promising explanation for the origin of degree
correlation observed in real networks of various types.

It will be of great interest to discuss the effect of node deleting
on degree correlation. In Fig.~\ref{r}, we give the assortativity
coefficient $r$ as a function of network size $N$, for different
$P_a$ in our model, for $m=5$. As one sees from Fig.~\ref{r}, for
each value of $P_a$, after a transitory period with finite-size
effect, each $r$ of networks tends to reach a steady value. When
$P_a=1$, $r\rightarrow0$ as $N$ becomes large. This result indicates
that networks in the BA model are uncorrelated, in agreement with
results obtained in previous studies \cite{30,33}. When $P_a<1$,
nontrivial negative degree correlations spontaneously develop as
networks evolve. One can see from Fig.~\ref{r} that the steady value
of $r$ in the model decreases with the decreasing $P_a$. In
particular, when $P_a\leqslant 0.6$, the value of $r$ is about
$-0.1$. These results indicate that node deleting leads to
disassortative mixing by degree in evolving networks. To make such
relation more clear, in Fig.~\ref{r-m}, we plot $r$ of networks in
our model as a function of $P_a$, for different $m$. As the
Fig.~\ref{r} indicates, when the network size is larger than
$40000$, the assortativity coefficient $r$ is nearly stable. So all
results in Fig.~\ref{r-m} are obtained from networks with $N=40000$.
Fig.~\ref{r-m} gives us the same relation between $r$ and $P_a$
shown in Fig.~\ref{r}. What is more, it tells us that for a given
$P_a$, $r$ will increase with the increasing $m$. The increment gets
its maximum between $m=1$ and other values. We point out that this
is because when $m=1$, the network has been broke up into small
separate components (see the following section). We can also find
from Fig.~\ref{r-m} that the gap between different curves decreases
with the increasing $m$ and the curves tend to merge at large $m$.

Now we give some explanations to the above observations. In the BA
model, the network being uncorrelated is the result of a competition
between two factors: the growth and the preferential attachment
(PA). On the one hand, networks with pure growth is positively
correlated. This is because the older nodes, also tending to be
higher degree ones, have a higher probability of being connected to
one another, since they coexisted earlier. In Fig.~\ref{r-nopa}, we
compute the assortativity coefficient $r$ of a randomly growing
network, which grows by the growth rule of BA-type, while the newly
added nodes connect to \emph{randomly chosen} existing ones. As one
can see from Fig.~\ref{r-nopa} that pure growth leads to positive
$r$. On the other hand, the introduction of PA makes the connection
between nodes tend to be negatively correlated, since newly added
nodes (usually low degree ones) prefer to connect to highly
connected ones. Then degree correlation characteristic of the BA
model is determined by this two factors. In Fig.~\ref{aknnba}, we
plot the average degree of the nearest neighbor $<k>_{nn}$ as a
function of $k$ in the BA model. It is found that nodes with large
$k$ show no obvious biases in their connections. But there is a
short disassortative mixing region when $k$ is relatively small
(also reported in Ref. \cite{a6}, see Fig.1a therein). Such
phenomenon can be explained by the effect of these two factor:
Growth together with PA makes nodes with large $k$ equally connect
to both large and small degree nodes, and the latter makes nodes
with small degree be disassortatively connected. Now, we introduce
node-deletion. According to Eq.~(\ref{eq:4}), depression of the
growth of large-degree nodes also decreases the connections between
them, therefore makes the correlation negative. We also investigate
the effect of node deleting on the $r$ of the randomly growing
network, and obtained similar results. As one sees from
Fig.~\ref{r-nopa}, depression of connections between higher degree
nodes causes the network less positively correlated, and with
stronger node-deletion, negatively correlated. Finally, with regard
to the effect of $m$ in this relation (Fig.~\ref{r-m}), larger $m$
means more edges are established according to the PA probability
Eq.~(\ref{eq:1}). We conjecture that the orderliness of newly added
nodes connecting to large degree nodes will be weakened by the
increasing randomness as $m$ becomes larger, thus leading to a less
negative correlation. Such randomness can not always increase and,
as we see from Fig.~\ref{r-m}, for large $m$, e.g., $m\geq14$, the
curves tend to merge together.

\section{\label{sec:s}SIZE OF GIANT COMPONENT}

\begin{figure}
\includegraphics[scale=0.8]{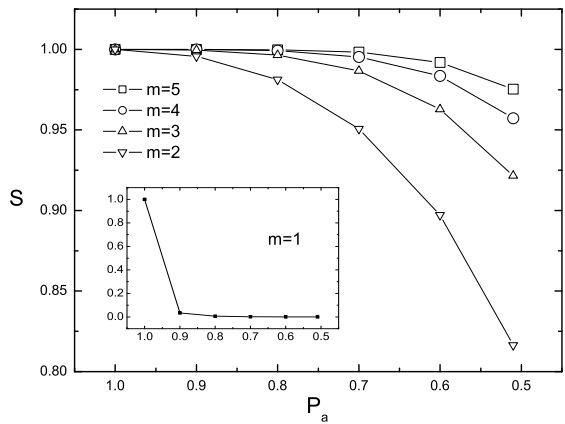}
\caption{\label{S} The relative size of the largest component $S$ as
a function of $P_a$ for $m=2,3,4,5$. Inset gives the same curve for
$m=1$. In the simulations, $N=100000$. All results are based on $10$
independent realizations.}
\end{figure}

In a network, a set of connected nodes forms a component. If the
relative size of the largest component $S$ in a network approaches a
nonzero value when the network is grown to infinite size, this
component is called the giant component of the network
\cite{1,a1,2}. In most previously studied growing models
\cite{1,a1,2}, due to the BA-type growth rule they adopted, there is
only one huge component in the network, i.e., $S\equiv1$. In this
extreme case the network gains a perfect connectedness. The opposite
case of $S=1$ is the extreme of $S=0$, in which case the network,
made up of small components, exhibits no connectedness. Experiments
indicate that some real networks seem to lie in somewhere between
these two extreme: they contain a giant component as well as many
separate components \cite{a1,2,36,37}. For example, According to
Ref.\cite{36}, in May of 1999, the entire WWW, containing
$203\times10^{6}$ pages, consisted of a giant component of
$186\times10^{6}$ pages and the disconnected components (DC) of
about $17\times10^{6}$ pages. In general, the introduction of node
deletion in our model will cause the emergence of separate
components even isolated nodes in the network. What we interest here
is the connectedness of the network. In Fig.~\ref{S} we plot the
relative size of the largest component $S$ in the model, as a
function of $P_a$, for $m=2,3,4,5$, where $m$ is the number of edges
generated with the adding of a new node. One sees from Fig.~\ref{S}
that for any $0.5<P_a\leq1$, a giant component can be observed in
the model if $m>1$. In addition, for the same $P_a$, $S$ increase as
the increase of $m$. While when $m=1$, the network is found to be
broke up into separate components if $P_a<1$. For example, when
$P_a=0.9$, $S$ of the network with $N=100000$ rapidly drops to
$0.034$. Inset of Fig.~\ref{S} gives the $S$ Vs $P_a$ curve for
$m=1$. These results indicate that node deleting does not destroy
the connectedness of a growing network so long as the increasing
rate of edges is not excessively small.

\section{\label{sec:l}AVERAGE DISTANCE BETWEEN NODES}
\begin{figure}
\includegraphics[scale=0.8]{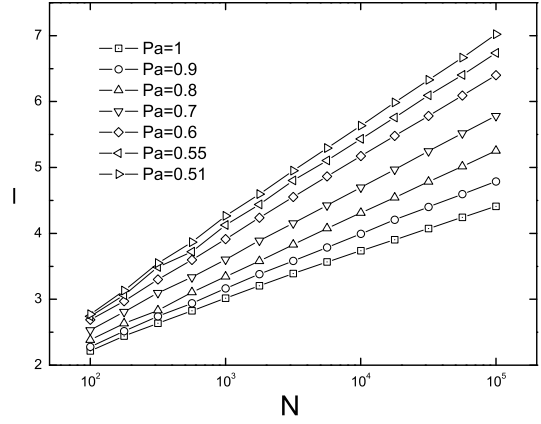}
\caption{\label{l} Average distance $L$ of the giant component in
the network as a function of network size $N$, for different $P_a$
in the model. The chose of some parameters: $m_0=m=5$. These curves
are results of $10$ independent realizations.}
\end{figure}

Now we study the effect of node deletion on networks' average
distance $L$ between nodes. Here the distance between any two nodes
is defined as the number of edges along the shortest path connecting
them. It has been revealed that, despite their often large size,
most real networks present a relatively short $L$, showing the
so-called small-world effect \cite{1,a1,2,3}. Such an effect has a
more precise meaning: networks are said to show the small-world
effect if the value of $L$ scales logarithmically or slower with
network size for fixed mean degree. This logarithmic scaling can be
proved for a variety of network models \cite{1,a1,2}. As we have
demonstrated in Section~\ref{sec:s}, node deleting does not destroy
the connectedness of the network in our model for any $m>1$, since
there is always a giant component exists. Here in our simulation, we
calculate $L$ of the giant component of the network in our model
using the ¡°burning algorithm¡± \cite{2}. In Fig.~\ref{l}, we plot
$L$ as a function of network size $N$, for different $P_a$ in our
model. As one can see from the figure, for any $0.5<P_a\leq1$, a
logarithmic scaling $L\sim\ln N$ is obtained, while the proportional
coefficient increases with the decrease of $P_a$. Furthermore, for a
given $N$, $L$ increases with the decrease of $P_a$. These results
tell us that node deleting will weaken but not eliminate the
small-world effect of a growing network.

\section{\label{sec:c}CLUSTERING}
\begin{figure}
\includegraphics[scale=0.8]{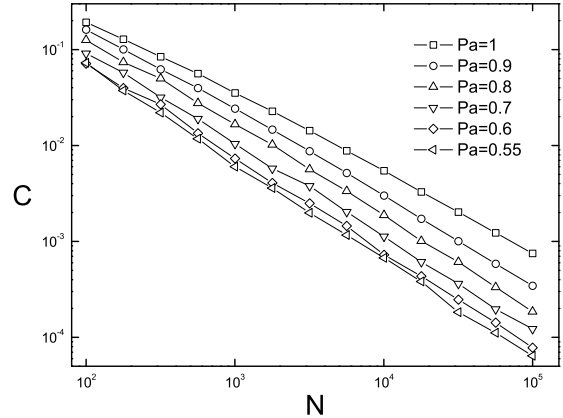}
\caption{\label{clustering} Cluster coefficient $C$ of the giant
component in the network as a function of network size $N$, for
different $P_a$. In the simulation we set $m_0=m=5$. These curves
are results of $10$ independent realizations.}
\end{figure}

Finally, we investigate the effect of node deletion on network's
cluster coefficient $C$, which is defined as the average probability
that two nodes connected to a same other node are also connected.
For a selected node $i$ with degree $k_i$ in the network, if there
are $E_i$ edges among its $k_i$ nearest neighbors, the cluster
coefficient $C_i$ of node $i$ is defined as
\begin{eqnarray}
C_i=\frac{2E_i}{k_i\left(k_i+1\right)}.
\end{eqnarray}
Then the clustering coefficient of the whole network is the average
of all individual $C_i$. In Fig.~\ref{clustering}, we plot $C$ of
the giant component in the network as a function of network size
$N$, for different $P_a$. As one sees from Fig.~\ref{clustering},
for each $P_a$, the clustering coefficient $C$ of our model
decreases with the network size, following approximately a power law
form. Such size-dependent property of $C$ is shared by many growing
network model \cite{1,a1,2}. Moreover, as Fig.~\ref{clustering}
shows, for the same network-size $N$, $C$ decreases as $P_a$
decreases. The results of Fig.~\ref{clustering} indicate that node
deleting weakens network's clustering.

\section{\label{sec:conclusion}CONCLUSION}
In summary, we have introduced a new type of network growth rule
which comprises of adding and deleting of nodes, and proposed an
evolving network model to investigate effects of node deleting on
network structure. It has been found that, with the introduction of
node deleting, network structure was significantly transformed. In
particular, degree distribution of the network undergoes a
transition from scale-free to exponential forms as the intensity of
node deleting increased. At the same time, nontrivial disassortative
degree correlation spontaneously develops as a natural result of
network evolution in the model. We also have demonstrated that node
deleting introduced in our model does not destroy the connectedness
of a growing network so long as the increasing rate of edge is not
excessively small. In addition, it has been observed that node
deleting will weaken but not eliminate the small-world effect of a
growing network. Finally, we have found that generally node deleting
will decrease the clustering coefficient in a network. These
nontrivial effects justify further studies of the effect of node
deleting on network function \cite{2}, which include topics such as
percolation, information and disease transportation, error and
attack tolerance, and so on.

\begin{acknowledgments}
The authors thank Doc. Ke Hu for useful discussions. This work is
supported by the National Natural Science Foundation of China, Grant
No. 10647132, and Natural Science Foundation of Hunan Province,
China, Grant No. 00JJY6008.
\end{acknowledgments}

\appendix*

\section{THE CALCULATION OF $S(T)$}
To get $S(t)$, we multiply both sides of Eq.~(\ref{eq:5}) by
$D'(i',t)$ and sum up all $i'$ between $0$ and $t$:
\begin{equation}
\sum_{i^{'}}\frac{\partial k(i',t)}{\partial t}D'(i',t)
=P_a(m-1)-\frac{1-P_a}{(2P_a-1)t}S(t)+1.
\label{eq:A1}
\end{equation}
To get the above equation we have used the definition of $S(t)$
[Eq.~(\ref{eq:6})] and the following equation:
\begin{eqnarray}
\sum_{i^{'}}D'(i',t)=\int_{0}^{t}D(i,t)di.
\end{eqnarray}
The left-hand side of Eq.~(\ref{eq:A1}) can be simplified as:
\begin{widetext}
\begin{eqnarray*}
\sum_{i^{'}}\frac{\partial
\left\{\left[k(i',t)+1\right]D'(i',t)\right\}}{\partial
t}-\sum_{i^{'}}\left[k(i',t)+1\right]\frac{\partial
D'(i',t)}{\partial t}\\=\frac{\partial }{\partial
t}\left\{\sum_{i^{'}}\left[k(i',t)+1\right]D'(i',t)\right\}-\left[k(t,t)+1\right]D(t,t)
\\-\sum_{i^{'}}\left[k(i',t)+1\right]D'(i',t)\frac{P_a-1}{(2P_a-1)t}.
\end{eqnarray*}
\end{widetext}
Substituting the above expression in Eq.~(\ref{eq:A1}), and noting
that $k(t,t)=m$ and $D(t,t)=P_a$, we get
\begin{eqnarray*}
\frac{\partial S(t)}{\partial
t}=\frac{2(P_a-1)}{(2P_a-1)t}S(t)+2P_am+1.
\end{eqnarray*}
The solution to the above equation is
\begin{eqnarray*}
S(t)=\left(2P_a-1\right)\left(2P_am+1\right)t.
\end{eqnarray*}

\end{document}